\documentclass[12pt,a4paper]{paper}

\usepackage[square,numbers]{natbib}
\usepackage{graphicx}
\usepackage{array}
 \usepackage{array,multirow}
  \usepackage{amsmath}
\usepackage{amsfonts}
\usepackage{xcolor}
\usepackage{booktabs}
\usepackage{hyperref}

\usepackage[utf8]{inputenc}

\newcommand{\rev}[1]{#1}

\begin{document}

%%%% Article title to be placed here
\title{Fusing electrical and elasticity imaging}

\author{%%%% Author details
Andreas Hauptmann$^{1,2}$, Danny Smyl$^{3}$ \\
%%%%%%%%% Insert author address here
\small$^{1}$University of Oulu, Research Unit of Mathematical Sciences, Finland\\
\small$^{2}$University College London, Department of Computer Science, UK\\
\small$^{3}$University of Sheffield, Department of Civil and Structural Engineering, UK  
}

\maketitle

%%%% Abstract text to be placed here %%%%%%%%%%%%
\begin{abstract}
 Electrical and elasticity imaging are promising modalities for a suite of different applications including medical tomography, non-destructive testing, and structural health monitoring.
    These emerging modalities are capable of providing remote, non-invasive, and low cost opportunities.
    Unfortunately, both modalities are severely ill-posed nonlinear inverse problems, susceptive to noise and modelling errors. 
    Nevertheless, the ability to incorporate compli-mentary data sets obtained simultaneously offers mutually-beneficial information.
    By fusing electrical and elastic modalities as a joint problem we are afforded the possibility to stabilise the inversion process via the utilisation of auxiliary information from both modalities as well as joint structural operators.
    In this study, we will discuss a possible approach to combine electrical and elasticity imaging in a joint reconstruction problem giving rise to novel multi-modality applications for use in both medical and structural engineering.
\end{abstract}
%%%%%%%%%%%%%%%%%%%%%%%%%%%

%%%%%%%%%% Insert the texts which can accomdate on firstpage in the tag "fmtext" %%%%%

\section{Introduction}
%\begin{itemize}
%    \item Motivation and applications
%    \item Why are we looking at this
%    \item What problem does it solve?
%\end{itemize}
The ability to quantitatively estimate and visualise structures and spatially-distributed parameter fields within an object is of central importance in a number of engineering and science applications.
This is especially true in the areas of medical imaging and non-destructive testing, where the accurate spatiotemporal estimation of anomalies can have significant implications on the future health and condition of the imaged target.
In the case of non-destructive testing, accurate localisation of deleterious structural anomalies can help in facilitating structural remediation before failure or impediment on user safety \cite{Farrar2007}.

In medical imaging, detection of tumours, haemorrhages, and other anomalies is essential for diagnostics and can potentially save lives \cite{bushberg2011essential}.
Among the broad spectrum of imaging modalities used in non-destructive testing (NDT) and medical imaging, MRI \cite{bogaert2012clinical}, X-ray tomography \cite{maire2014quantitative}, and acoustic tomography \cite{behnia2014advanced}
%, and photoacoustic imaging \cite{cox2012quantitative} 
have been the source of significant industrial and academic research in recent years.
In addition to mainstream modalities, Electrical Impedance Tomography (EIT) \cite{mueller2020d,shin2020second,smyl2020electrical,hamilton2019beltrami,Tallman2015,hallaji2014electrical,Tallman2014,Gomez-Laberge2012,Tawil2011,Brown2003} and Quasi-Static Elasticity Imaging (QSEI) \cite{Smyl2019SHM,smyl2018coupled,goenezen2011,gokhale2008,oberai2003} have emerged as promising quantitative approaches to imaging and characterising biological and structural targets.
Both imaging approaches are similar, in that, they are used to interpret internal structures from distributions (or changes in distributions) of material properties.

In EIT, internal structures (such as lung boundaries \cite{liu2019moving} or cracks in structural testing \cite{smyl2018detection,hauptmann2018revealing,ikehata2016enclosure}) and their temporal evolution are monitored and interpreted via conductivity distributions reconstructed from boundary voltage measurements.
On the other hand, QSEI aims to reconstruct the distribution of the elastic modulus from distributed displacements computed using, e.g., digital image correlation (DIC) \cite{smyl2018coupled} or pixel changes in ultrasound images \cite{oberai2003}.
However, despite their similar aim in reconstructing constitutive fields representative of the underlying structural characteristics, the physics and inverse determination of these fields using EIT and QSEI are drastically different.

Fundamentally, EIT is a diffusive imaging modality, where diffusive electric currents are injected into a target and the corresponding boundary voltage measurements are non-linearly related to the internal conductivity distribution.
Conversely, QSEI displacement measurements are "full field," i.e. discrete measurements (usually evenly) covering entire target.
Based on these unique differences, each modality has seperate strengths and weaknesses -- for example, QSEI is more sensitive to localised changes far from the boundary and EIT is more sensitive to changes near the electrodes\rev{, thus offering a natural synergy to improve reconstruction quality for both modalities.}

These differences, however, offer an excellent opportunity to leverage the strengths of each modality in order to reconstruct images at higher fidelity than would be possible using a single modality.
One approach to fusing the former modalities is the use of joint inversion, which is classically accomplished using a joint structural operator acting to iteratively penalise discrepancies between reconstructions \cite{haber1997joint}.
In the case of joint EIT-QSEI imaging, this was first proposed computationally in \cite{Smyl2019SHM} where a binary penalty was applied.
One central challenge in adopting such an approach is that the binary joint penalty term in the functional is not differentiable and is thus difficult to incorporate in regularising and formulating the minimiser.
As such, significant spatial information is not available during reconstruction, thereby significantly reducing the potential for accurate spatial imaging of the elasticity and conductivity fields.

In this work, we \rev{discuss how to effectively combine the structural information of both modalities by} reformulating the joint EIT-QSEI problem utilising a joint total variation (JTV) prior. \rev{This joint prior enables a suitable, effective regularisation and appropriate penalisation of} each modality in order to improve joint reconstruction quality and fidelity.
This will be accomplished by first reviewing the forward problems for each modality and \rev{formulating a fused JTV framework that is suitable for the task at hand}.
Following, results from a computational investigation will be reported and analysed\rev{, demonstrating increased image quality and robustness to noise.}
Lastly, discussion and outlook for fused EIT-QSEI imaging will be provided.

\section{Joint electrical and elasticity imaging}
The reconstruction task for either EIT or QSEI is ill-posed and hence suitable regularisation techniques need be devised. In this work we consider a variational approach, where we aim to obtain reconstructions as the minimiser of a cost function offering the possibility to flexibly incorporate suitable prior knowledge into the recovery task as regulariser, which can act on both quantities separately and/or jointly. In particular, in the following we will shortly discuss the reconstruction task for both sub-problems\rev{, how both complement each other} and then introduce the \rev{specially designed joint formulation based on the JTV framework}.

\subsection{Electrical impedance tomography}
We will consider here a two-dimensional bounded domain $\Omega\subset\mathbb{R}^2$, but note that an extension to three dimensions is straight-forward. For conducting electrical measurements we place a set of $M$ disjoint electrodes on the boundary $E_m\subset\partial\Omega$ for $1\leq m \leq M$. For the measurement process, we apply currents $I_m$ at the boundary and measure the resulting voltage at each electrode $U_m$, given a conductivity distribution inside the domain $\gamma(x)$. The full model is then given by the established complete electrode model \cite{somersalo92}:
\begin{equation}
\label{eqn:CEM}
\left\{
\begin{array}{rcll}
\vspace{0.2 cm}
\nabla\cdot(\gamma(x)\nabla u(x)) &=& 0 &\mbox{ for } x \in \Omega,\\ 
\vspace{0.2 cm}
u(x)+z_m \gamma(x) \frac{\partial u(x)}{\partial \bar{n}} &=& U_m &\mbox{ for } x\in E_m,\, 1\leq m \leq M  \\
\vspace{0.2 cm}
\int_{E_m} \gamma(x) \frac{\partial u(x)}{\partial \bar{n}} \mathrm{d}S  &=& I_m &\mbox{ for } 1\leq m \leq M \\
\gamma(x) \frac{\partial u(x)}{\partial \bar{n}} &=& 0 &\mbox{ on }\partial\Omega\backslash \bigcup_{m=1}^M E_m
\end{array}
\right.
\end{equation}
Here $z_m$ denotes the contact impedance at each electrode and $\bar{n}$ the outer unit normal on the boundary. We note that due to conservation of charge all injected currents sum to zero, similarly a ground level for the voltage potential can be fixed by assuming that the measured voltages sum to zero as well. The forward model is then given by the mapping from conductivity $\gamma$ to a set of measured voltages $V$, we write $\gamma \mapsto U_\mathrm{EIT}(\gamma)$.

Given the set of measured voltages $V$, we can compute a reconstruction of the conductivity $\gamma$ as the minimiser of a penalty functional 
\begin{equation}\label{eqn:penaltyFuncEIT}
    \Psi_{\mathrm{EIT}}(\gamma) = || V - U_\mathrm{EIT}(\gamma)||^2_2 + \alpha\mathcal{R}(\gamma),
\end{equation}
where $\mathcal{R}$ denotes a suitable regulariser and $\alpha>0$ is a weighting parameter balancing the data fidelity term and the influence of prior information contained in the regulariser.

Specifically, the regulariser does not only stabilise the inversion process, but also offers the possibility to flexibly incorporate prior information into the reconstruction task. Such prior information can come in large variety, from simple assumptions on continuity \cite{vauhkonen1998tikhonov} or piece-wise constant reconstructions, incorporated as a total variation penalty \cite{gonzalez2017}, to more informative priors using structural information  \cite{kolehmainen2019incorporating} and geometric assumptions \cite{liu2020shape}. In particular, as the reconstruction task in EIT is non-linear and highly ill-posed such prior information is essential to improve reconstruction quality. This positive effect can be even clearly demonstrated for direct reconstructions with the D-bar method \cite{alsaker2016d,alsaker2017direct,alsaker2019dynamic} and recent advances in utilising deep learning techniques to infer prior  information from large datasets \cite{hamilton2019beltrami, hamilton2018deep}.  

In this work, rather than using specially designed priors, we consider the possibility to supply auxiliary information from a secondary modality providing parallel measurements and thus improving the reconstruction quality.
\rev{In particular, a complementary modality would primarily need to compensate for the diffusive nature of EIT. That means, in EIT measurements are only obtained at the boundary from diffusive electric currents and thus loss of contrast as well as distortions are stronger within the target, but a higher accuracy at the boundary can be obtained. Consequently, we will discuss next the synergy with a full-field modality to improve reconstruction quality over the whole domain.}

\subsection{Quasi-static elasticity imaging}
In contrast to boundary measurements, in QSEI the inverse problem is to compute the distributed elasticity modulus $E$ from a measured displacement field $u_m$ inside the domain.
To be more precise, the elastic forward model is defined herein for the same two-dimensional geometry $\Omega\subset\mathbb{R}^2$ with boundary $\partial \Omega$ as above.
On part of the boundary $\partial \Omega_{1}$ a prescribed boundary traction $f(x):\partial\Omega_1\to\mathbb{R}$ is applied, referred to as imposed boundary conditions. 
Concurrently, we denote the (unknown) displacements on $\partial \Omega_{2}$ by $\hat{u}$.
Neglecting body forces, the isotropic QSEI forward problem is then written as follows

% \begin{equation}
% \frac{\partial \sigma}{\partial x} = 0, \textbf{x} \in \Omega
% \label{QSEI1}
% \end{equation}

% \begin{equation}
% f(\textbf{x})  \equiv \sigma \bar{n}  \equiv \hat{f}, \textbf{x} \in \partial \Omega_{1}
% \label{QSEI2}
% \end{equation}

% \begin{equation}
% u = \hat{u}, \in \partial \Omega_{2}
% \label{QSEI3}
% \end{equation}

\begin{equation}
\label{eqn:QSEIF}
\left\{
\begin{array}{rcll}
\vspace{0.2 cm}
\nabla \cdot \sigma(x)  &=& 0, &\mbox{ for }  x \in \Omega \\
\vspace{0.2 cm}
   \sigma(x) \bar{n}(x)  &=& f(x), &\mbox{ for } x \in \partial \Omega_{1} \\
 u &=& \hat{u},  &\mbox{ on } \partial \Omega_{2}
\end{array}
\right.
%\label{QSEIF}
\end{equation}

\noindent where $\sigma$ is the  two-dimensional plane stress tensor, $\bar{n}$ the outer unit normal, and $\partial \Omega_{1} \cup \partial \Omega_{2} = \partial \Omega$. 
We then obtain a Finite Element discretisation by utilising the weak Galerkin formulation of Eq. \ref{eqn:QSEIF} \cite{surana2016} as follows.
From the discretised problem, we formulate a system of linear equations as a function of the elastic modulus $E$, which may be written as $\sum_{i=1}^n K_{ij}({E})u_j=F_i$ where $K_{ij}$ is often referred to as the stiffness matrix and $F_i$ is, correspondingly, a vector of (nodally) applied forces.
In the QSEI inverse problem, however, we aim to compute the displacement field $u$, therefore the following form is used $u_j = \sum_{i=1}^n K_{ij}^{-1}({E})F_i $ for $j=1,2$ and $n$ refers to the number of unknown displacements.

In particular, the recovery of $E$ from a measured displacement field $u_m$ is (as well) a non-linear inverse problem, i.e. the forward model $E\mapsto U_\mathrm{QSEI}(E)$ for the simulated displacement field is nonlinear in $E$.
Generally, one aims to minimise the following penalty functional to reconstruct $E$ from displacement measurements $u_m$

\begin{equation}\label{eqn:penaltyFuncQSEI}
    \Psi_{\mathrm{QSEI}}(\gamma) = || u_m - U_\mathrm{QSEI}(E)||^2_2 + \alpha\mathcal{R}(E),
\end{equation}

\noindent where $\mathcal{R}$ and $\alpha$ take the same meaning as described in the previous subsection.
As a whole, a number of inversion approaches have been used to solve the QSEI problem.
Such methods include, for example, efficient adjoint frameworks \cite{oberai2003}, Gauss-Newton based approaches \cite{bonnet2005inverse}, data-driven \cite{hoerig2018data}, and stacked methods in the case of anisotropy/orthotropy \cite{mei2019quantifying,smyl2018coupled}.
For a more comprehensive discussion on non-linear constitutive modelling and solving the QSEI problem, we refer the reader to a number of excellent works by Goenezen, Barbone, Oberai, and co-authors (e.g. \cite{goenezen2011, barbone2002,barbone2004elastic, oberai2003,oberai2003solution,oberai2004}).

{It is worth mentioning here what advantage QSEI may provide in the joint imaging framework. 
Since QSEI is a full-field imaging modality, it is highly sensitive to localized parameter changes in $\Omega$.
In contrast, EIT is a diffusive modality and is therefore less sensitive to localized parameter changes.
Further, the noted sensitivity decreases proportionally with the distance from the electrodes.
As such, it is surmised that including QSEI in the joint imaging framework may compensate for the lack of EIT sensitivity far from the electrodes and in return joint imaging may reduce noise sensitivity for both modalities -- a hypothesis to be confirmed in the results section.}

\subsection{Solving the joint problem}
Formulating a joint EIT-QSEI imaging framework aims to take advantage of two mutually-beneficial factors (a) the sensitivity of QSEI and EIT to changes far from and near to the boundaries, respectively and (b) the assumption of anomaly sharpness within the imaged target $\Omega$.
To capitalise on (a), full data sets from both EIT and QSEI will be utilised and for (b), a bespoke form of the joint total variation (JTV) \rev{will be formulated that is capable to penalise spatial smoothness, while benefiting from shared information and maintaining the correct quantitative values of both modalities.}
With these factors identified, instead of computing reconstructions for each problem separately in Eqs. \ref{eqn:penaltyFuncEIT} and \ref{eqn:penaltyFuncQSEI}, we aim to recover both reconstruction jointly by minimising the following joint functional
\begin{equation}
\Psi(E,\gamma) = ||V - U_\mathrm{EIT}(\gamma)||^2_2 + ||u_m - U_\mathrm{QSEI}(E)||^2_2 + \alpha J(\gamma, E) + \lambda||\gamma||_1 + \lambda||E||_1
\label{JointF}
\end{equation}

\noindent where $U$ (and the appropriate subscript) represent the non-linear forward models, $\alpha$ is the JTV weighting parameter, and $\lambda$ is the $L^1$ regularisation parameter. \rev{We here included the $L^1$-norm as regulariser to first stabilise the inversion process and secondly promote sparse crack like structures}. We note that the sum of the total variation and the $L^1$-norm in fact defines the norm for functions of bounded variation (BV), thus the penalty in Eq. \ref{JointF} can be considered as a BV-regularisation term.

\rev{Nevertheless, we need to carefully adjust the joint total variation regulariser to the problem at hand, since a primary} concern in the joint recovery for EIT-QSEI imaging is to preserve the quantitative values for both modalities, while benefiting from shared information on regularity and edge alignment. \rev{To achieve this, we first write} the JTV functional $J$ in the discrete form as follows 
\begin{equation}
J(\gamma, E) = \sum_{h=1}^{H} \sqrt{ |\nabla \gamma_h|^2 + |\kappa \nabla  E_h|^2 + \beta}
\label{JTVF}
\end{equation}

\noindent where $h$ refers to a given element in the discretisation, $H$ are the total number of degrees of freedom in the discretisation, $\nabla$ is the finite difference operator, $\kappa$ refers to a \rev{crucial} scaling parameter (to be discussed later), and $\beta$ is a small parameter needed to make $J$ differentiable.
More explicitly, we may write Eq. \ref{JTVF} in the full isotropic discretised form including directional derivatives in both spatial directions

% \begin{equation}
% J(\gamma, E) = \sum_{h=1}^{H} \sqrt{ (\nabla_x \gamma_h)^2 + (\nabla_y \gamma_h)^2 + (\kappa \nabla_x  E_h)^2 + (\kappa \nabla_y  E_h)^2 + \beta}.
% \label{JTVF2}
% \end{equation}

\begin{equation}
J(\gamma, E) = \sum_{h=1}^{H} \sqrt{ (\partial_{x_1} \gamma_h)^2 + (\partial_{x_2} \gamma_h)^2 + (\kappa \partial_{x_1}  E_h)^2 + (\kappa \partial_{x_2} E_h)^2 + \beta}.
\label{JTVF2}
\end{equation}

It is important to note the the JTV functional described in Eq. \ref{JTVF2} is familiar to joint TV functionals used in previous works (e.g. \cite{ehrhardt2016multicontrast,chen2013calibrationless,zhang2018pet}).
However, unlike similar works using either a level-set or global normalisation scheme (e.g., the optimisation parameter is scaled from 0 to 1), we choose to scale one of the joint parameters with respect to the other for simplicity and to prevent over regularisation of one modality \rev{while reducing necessary regularisation parameters.} 
To do this, the parameter $\kappa$ is used to scale $E$ proportionally to $\gamma$ for computing the regulariser.
For this, we choose the rational scaling of $\kappa = \frac{\gamma_\mathrm{exp}}{E_\mathrm{exp}}$, where $\gamma_\mathrm{exp}$ and $E_\mathrm{exp}$ are the expected (scalar) values computed as the best homogeneous estimates. {The homogeneous estimates are computed following \cite{Smyl2017e}.}
Finally to compute the reconstructions by minimising Eq. \ref{JointF}, a Gauss-Newton scheme equipped with a linesearch was adopted where iterations were terminated when the change in the cost functional was less than $10^{-2}$. 

\section{Computational results}

\subsection{Computational setup}
\begin{figure}[h]
\centering
\scalebox{0.675}{
\begin{tabular}{ccccc}
\includegraphics[width=35mm]{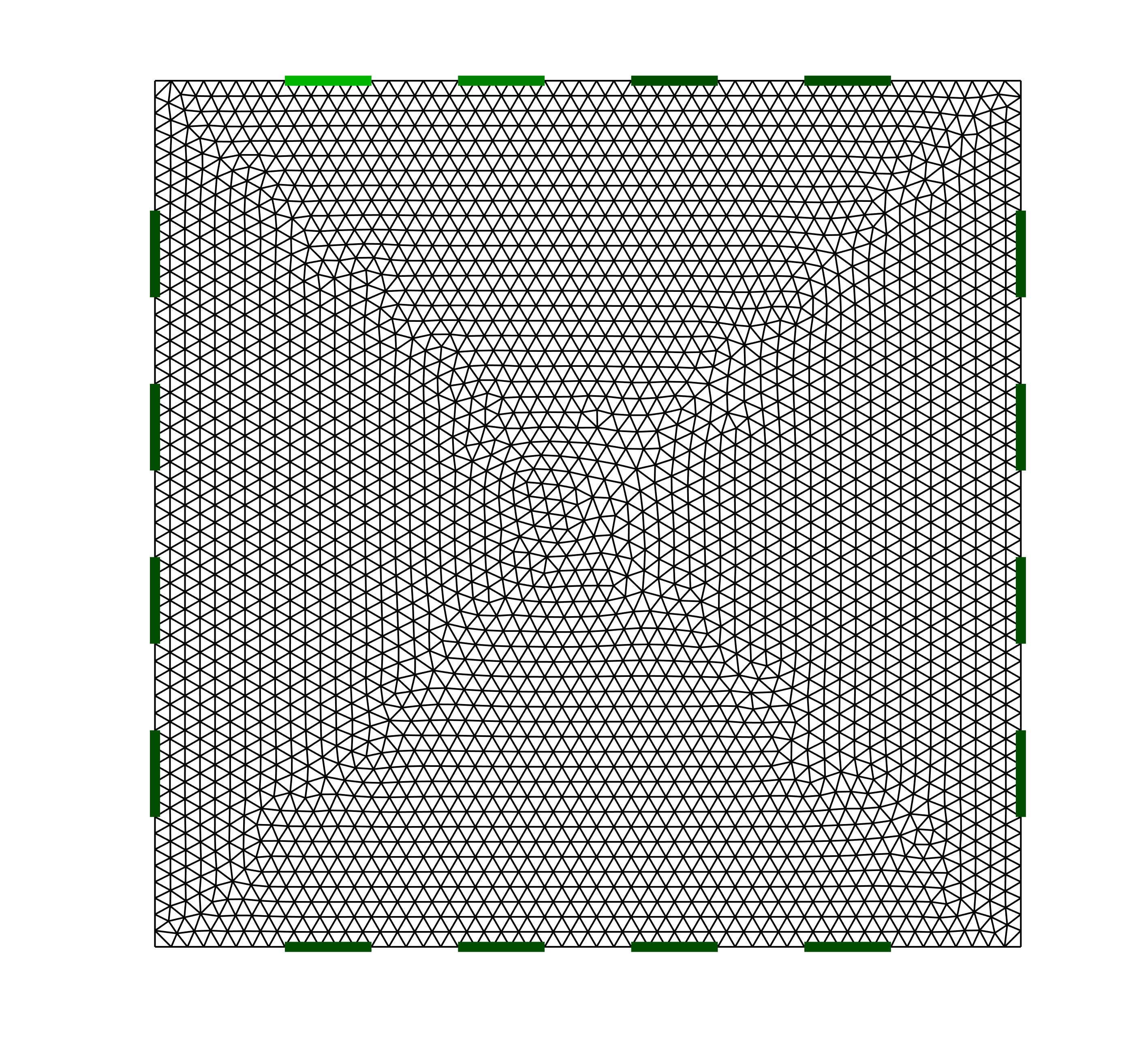} &  \includegraphics[width=35mm]{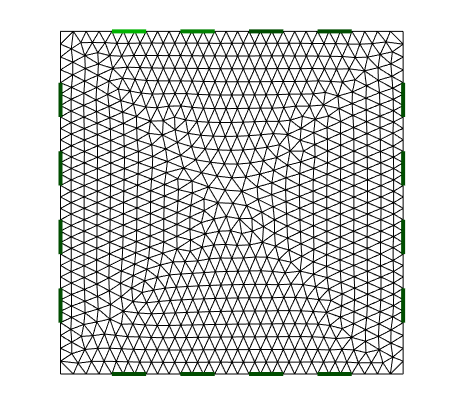} & \includegraphics[width=35mm]{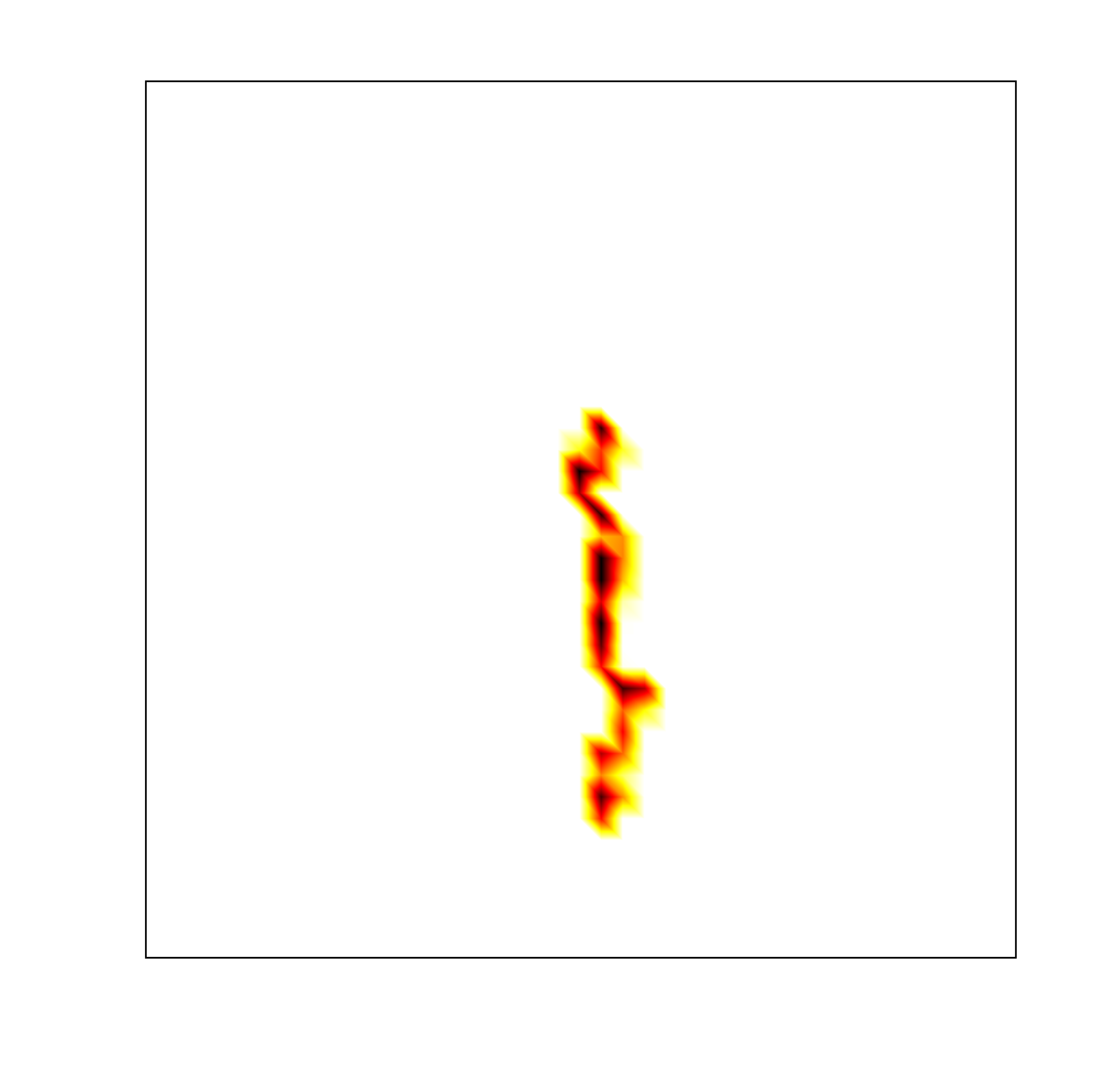} &
\includegraphics[width=35mm]{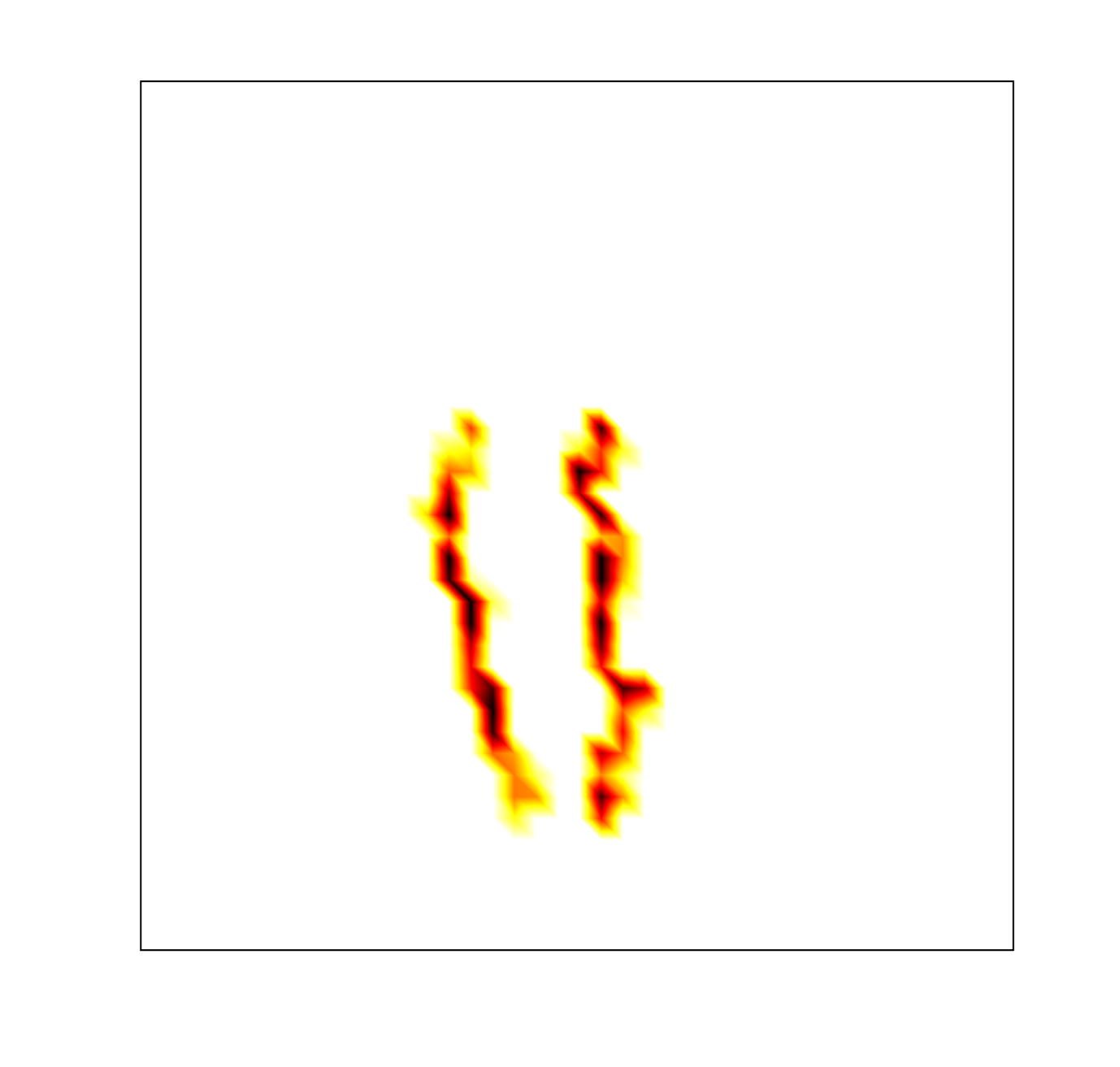} &
\includegraphics[width=35mm]{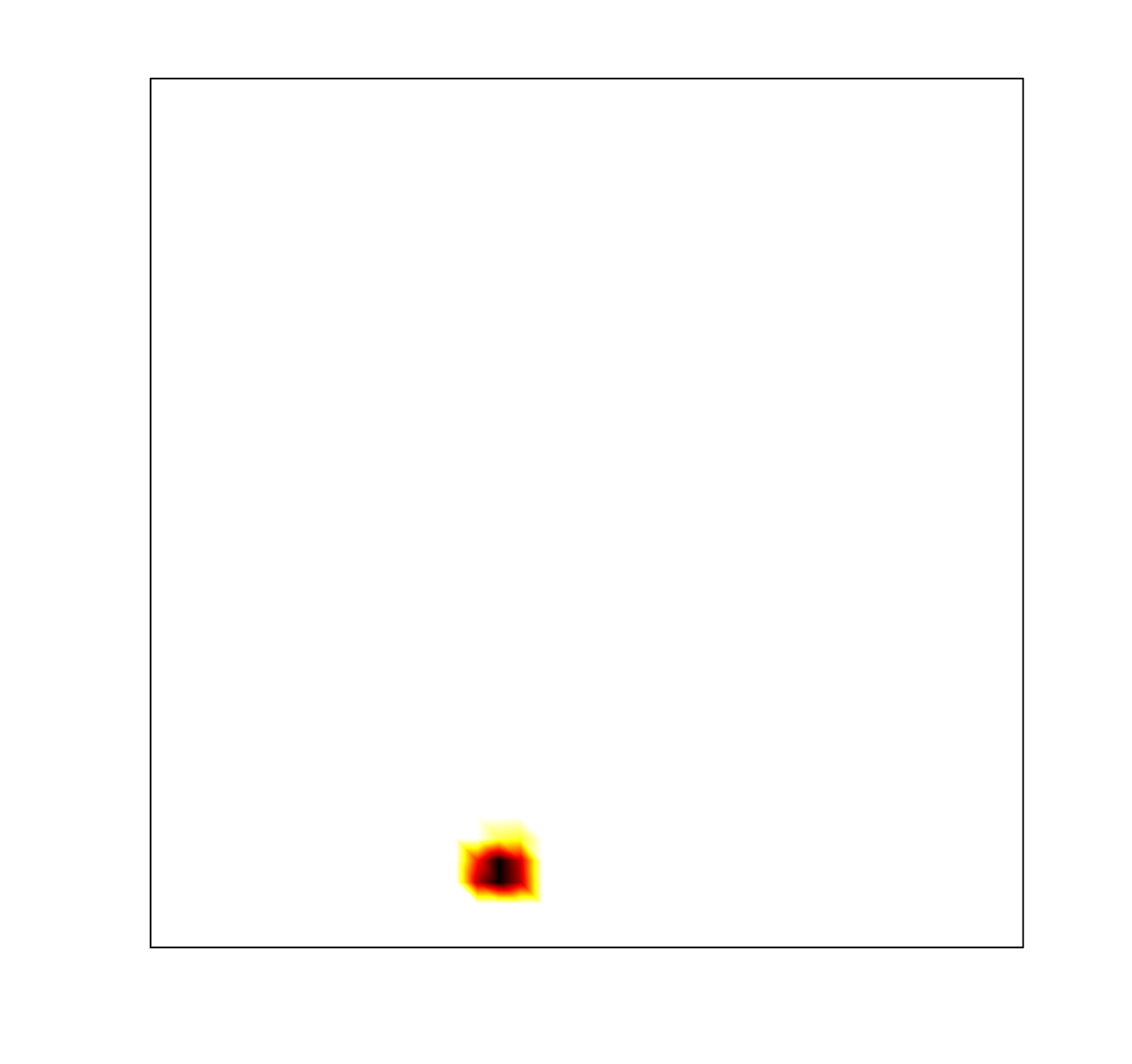} \\
(a) & (b) & (c) & (d) & (e)\\
\end{tabular}}
\caption{Computational information used in the numerical study: (a) fine mesh used for data generation consisting of 5730 elements, (b) inversion mesh consisting of 1556 elements, (c) the {single crack} target crack geometry, { (d) the double crack target geometry, and (e) a through hole (puncture damage).}.  The same meshes were used in data generation and inversion for both EIT and QSEI, where the background values were set to $E_\mathrm{background} = 200$ GPa and $\sigma_\mathrm{background} = 20$ S/m, consistent with structural steel (Poisson ratio, $\nu = 0.3$) and conductive silver sensing skin, respectively.}
\label{schem}
\end{figure}

In the following we will evaluate the effectiveness of the proposed joint reconstruction framework by comparing it to the respective single-modality reconstructions.
In doing this, we {first} simulate a realistic crack, shown in Fig. \ref{schem}(c), within a (left-end) clamped $1 \times 1$m square domain representative of a 5mm thick plate subjected to a 10kN/m distributed end load to be non-destructively evaluated using QSEI and EIT.
For this study, the crack was considered non-conductive ($\sigma \approx  0$) and non-elastic ($E \approx 0$), consistent with a structural through crack \cite{smyl2018detection}.
{Following, using the same crack characteristics, we simulate a double crack case as shown in Fig. \ref{schem}(d).
Lastly, an edge through hole simulating puncture or impact was generated as depicted in Fig. \ref{schem}(e).}
To prevent inverse crime, separate data generation and inversion meshes are used which are shown in Figs. \ref{schem}(a-b).
Note that the same data generation and coarse inversion discretisations are used for EIT and QSEI.

In generating the EIT data, 16 boundary electrodes were used in conjunction with a 1mA opposite electrode current injection and adjacent electrode measurement protocol.
Correspondingly, the electrode contact impedances were set to 0.01$\Omega \mathrm{cm}^2$.
This EIT measurement strategy resulted in 192 measurements.
Meanwhile, QSEI measurements were collected from the fine mesh nodes and interpolated to the coarse mesh nodes using linear interpolation, thus resulting in 1674 total displacement measurements (including $x$- and $y$- displacement fields).
After collecting EIT and QSEI measurements, 1\% noise standard deviation was added.

{All reconstructions were carried out on a quad-core Intel i7-7700HQ machine equipped with 16GB of RAM.
Generally, JTV computations required 20-30 iterations to reach stopping criteria with each iteration taking approximately 3-5 minutes (primarily in computing the Jacobians).
Single modality reconstructions each required approximately 10-50\% of the former computing time depending on the modality.}

%%%%%%%%%%%%%%%%%%%%%%%
\subsection{Inversion parameter selection}
In this study, both joint and single-modality inversion frameworks are tested and compared.
Importantly, two  parameters, $\alpha$ and $\lambda$, are used for JTV/TV and the $L^1$ regularisation term, respectively, in addition to the JTV/TV stabilisation parameter $\beta$. We emphasise here, that also the single-modality framework uses a TV and $L^1$ penalty for fair comparisons. 
In order to rationally select the regularisation parameters in the experiments, we first utilise the methodology proposed in \cite{gonzalez2017}, where the TV weighting parameter is selected using $\alpha = \frac{\ln (1-p/100)}{\sigma_\mathrm{exp}/d}$. Namely, $d$ is the element width and $p$ is the probability (\%) that the estimated parameter is predicted to lie within the expected range -- herein we use $p=99$.
Recalling that $\kappa$ is used to  normalise gradients of $E$, the expected range is accurately captured by $\sigma_\mathrm{exp}$ (since $\sigma$ estimates are to expected lie between $\sigma_\mathrm{exp}$ and zero).
For both $\lambda$ and $\beta$, empirical testing was used to select appropriate values, for which $\beta = 10^{-3}$ and $\lambda = 10^{-5}$ were used throughout the study.

%%%%%%%%%%%%%%%%%%%%%%
\subsection{Results}
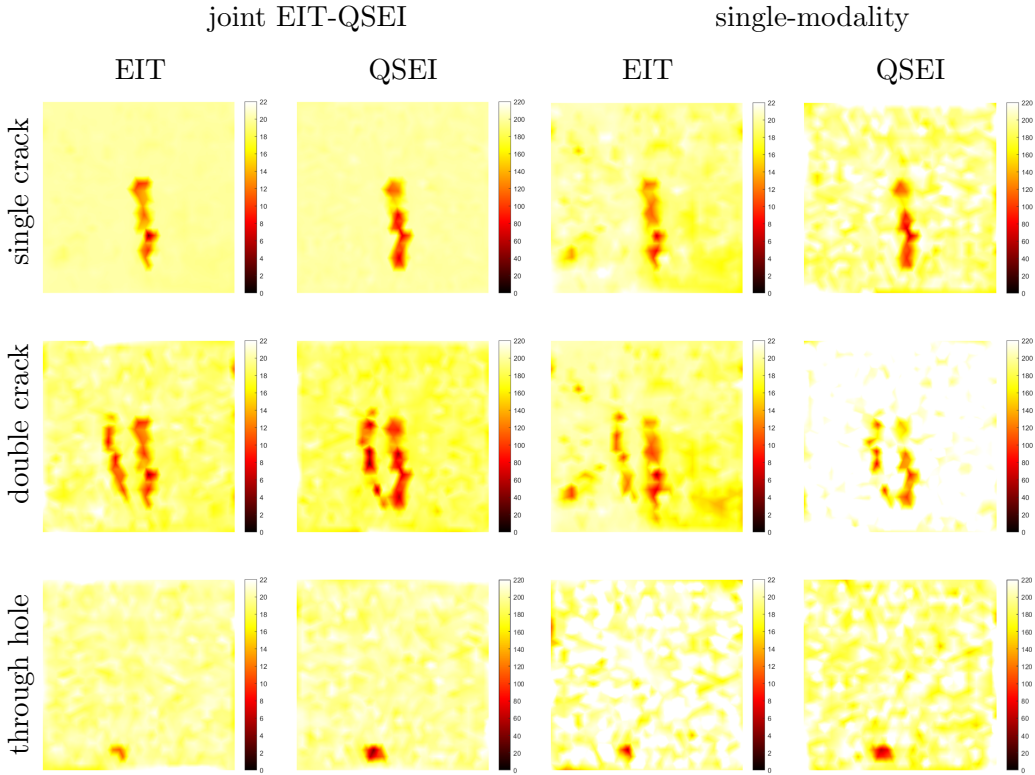
\begin{figure}[h!]
\centering
\centering
\small
\begin{picture}(400,300)

\put(10,180){\includegraphics[width=90pt]{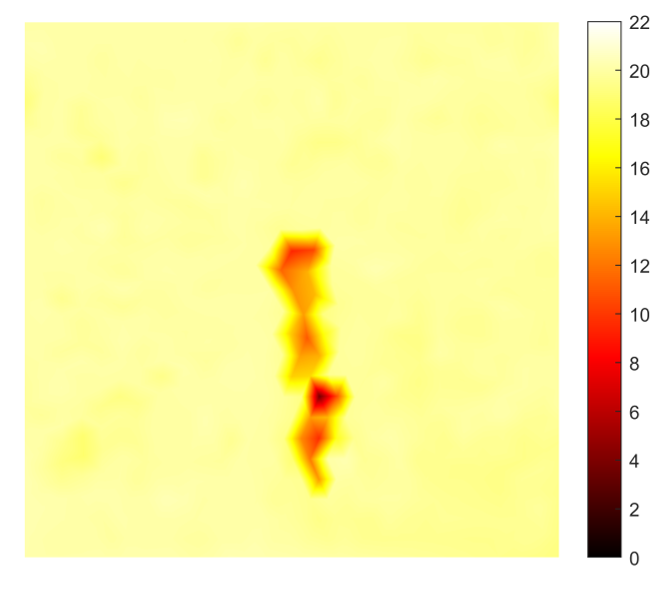}} 
\put(105,180){\includegraphics[width=90pt]{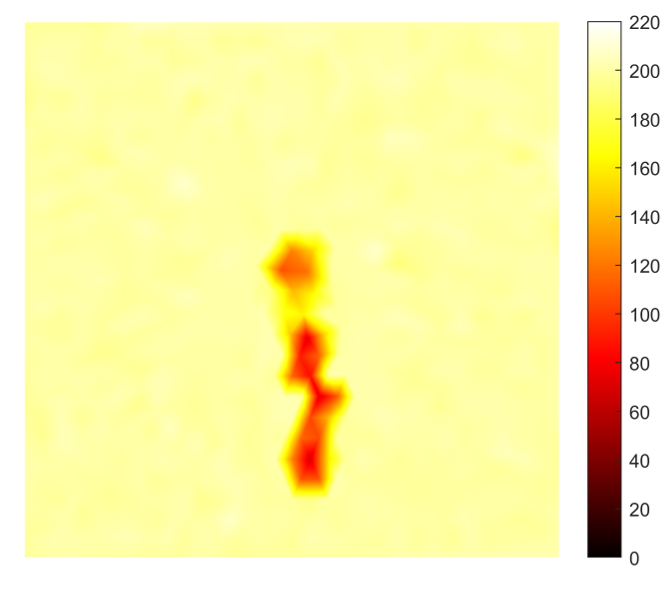}}
\put(200,180){\includegraphics[width=90pt]{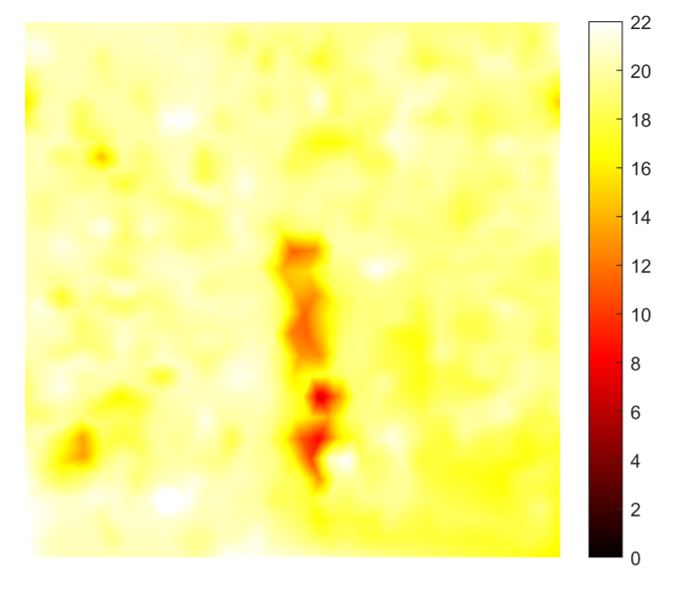}}  
\put(295,180){\includegraphics[width=90pt]{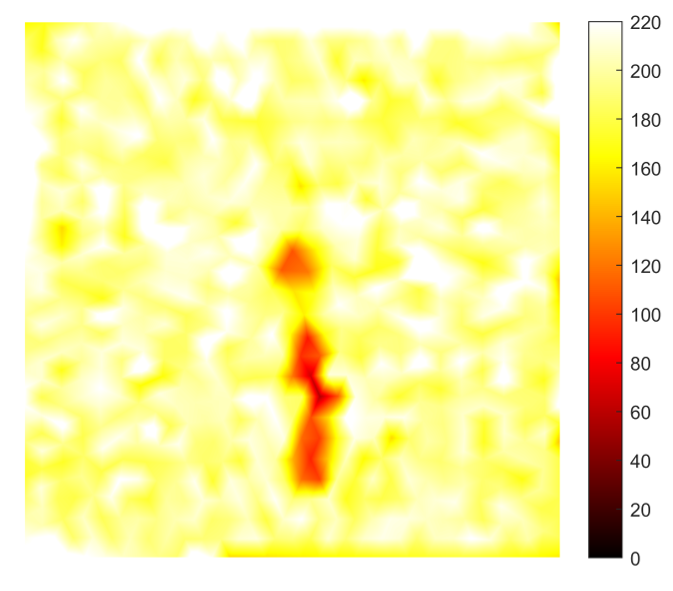} }

\put(10,90){\includegraphics[width=90pt]{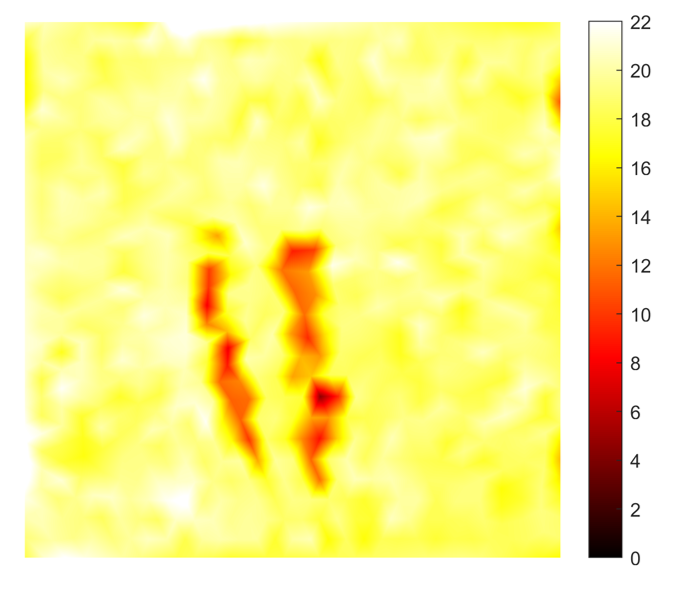}}
\put(105,90){\includegraphics[width=90pt]{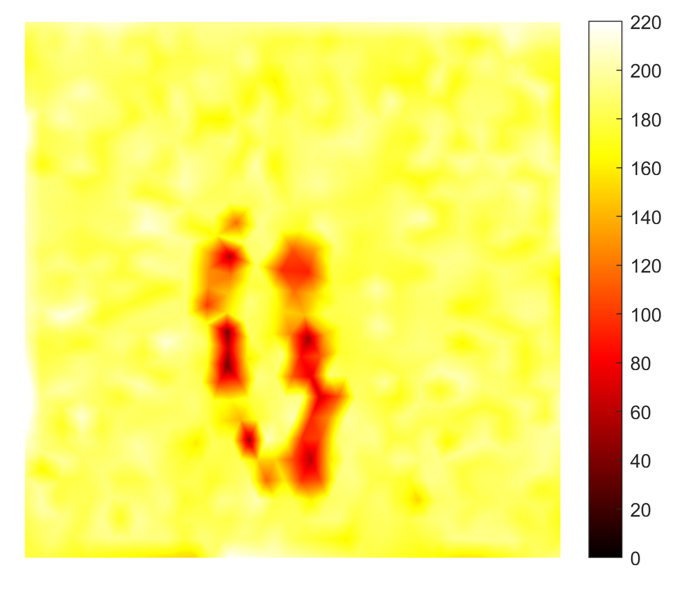}}

\put(200,90){\includegraphics[width=90pt]{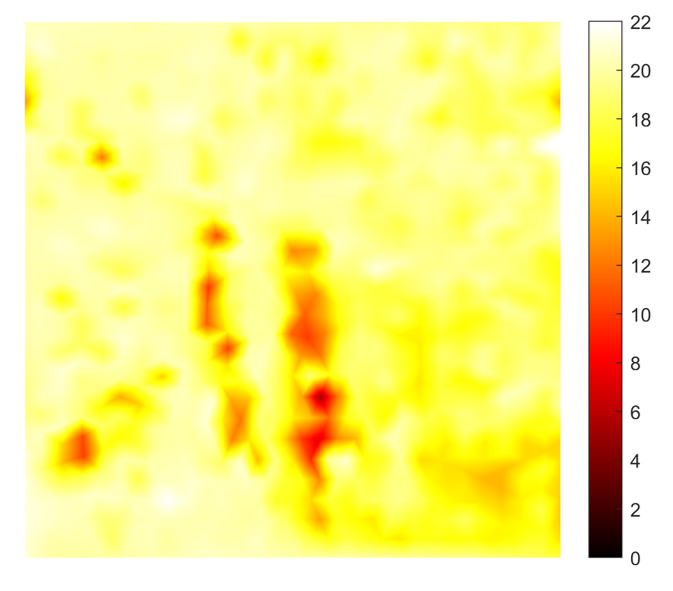}}
\put(295,90){\includegraphics[width=90pt]{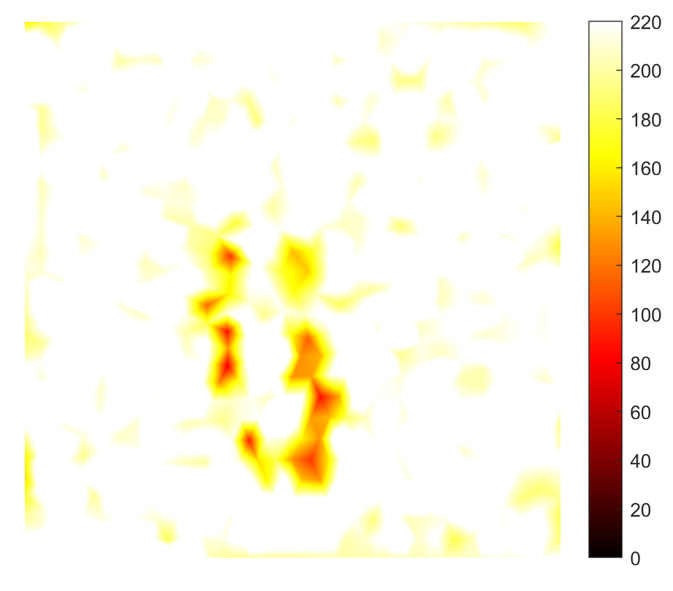}}

\put(10,0){\includegraphics[width=90pt]{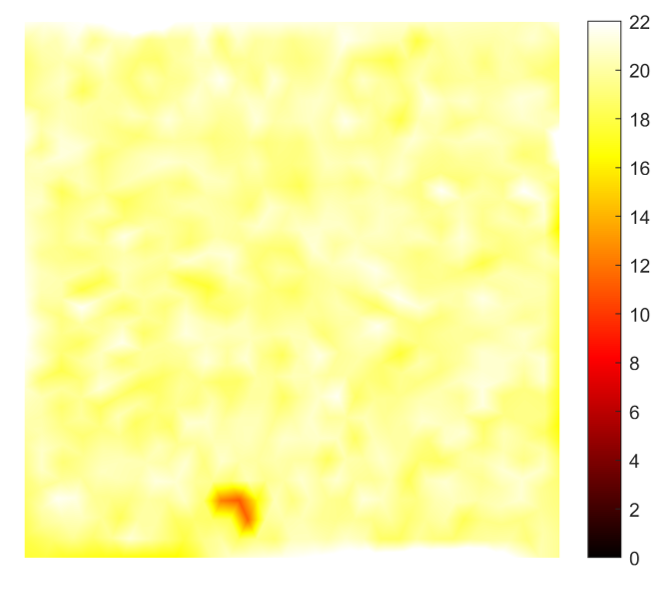}}
\put(105,0){\includegraphics[width=90pt]{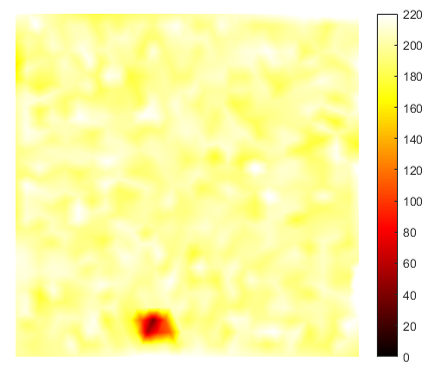}}

\put(200,0){\includegraphics[width=90pt]{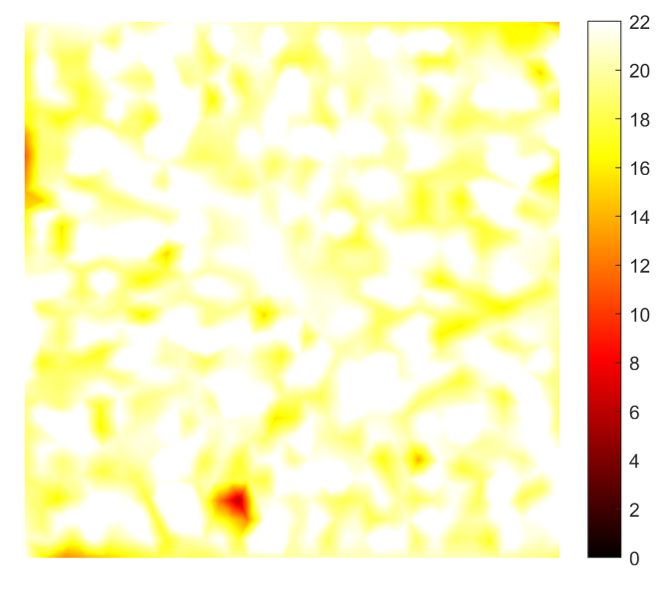}}
\put(295,0){\includegraphics[width=90pt]{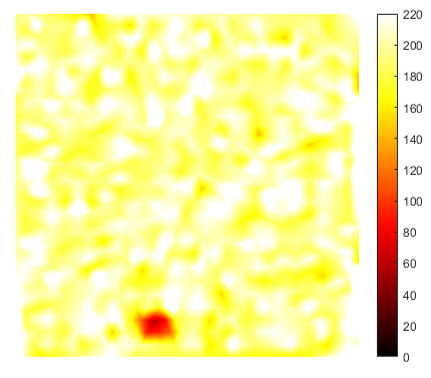}}

\put(40,265){EIT}
\put(135,265){QSEI}
\put(230,265){EIT}
\put(325,265){QSEI}

\put(0,10){\rotatebox{90}{through hole}}
\put(0,105){\rotatebox{90}{double crack}}
\put(0,195){\rotatebox{90}{single crack}}

\put(75,285){joint EIT-QSEI}
\put(265,285){single-modality}

%\textbf{EIT (mS/cm)} & \textbf{QSEI (GPa)} \\
\end{picture}
\caption{Comparison of obtained reconstructions of conductivity $\sigma$ (mS/cm) and elasticity field $E$ (GPa), with joint total variation and separate single-modality TV \rev{for three test cases}.}
\label{results}
\end{figure}

In this subsection, we report the reconstructions using JTV and single-modality (TV regularised) frameworks. 
These images are shown in Fig. \ref{results}, where {EIT and QSEI reconstructions are shown in the left and right columns, respectively.}
%the first row shows JTV reconstructions and the second row shows single-modality reconstructions.
As a whole, all reconstructions localise the crack(s) well, while the single-modality images {generally show more} artefacts in the background and notably less continuity along the crack length(s).
{The latter is particularly true in the case of the double crack, where the continuity of cracks is far less distinguishable when single modality reconstruction is used.}
These visual observations are quantitatively supported by computing the mean square error {(MSE, normalised with respect to the ground-truth for each modality), as reported in Table \ref{reconinfo}.
Indeed, image MSEs reported in Table \ref{reconinfo} are notable lower in cases where JTV is used.}
%where EIT and QSEI JTV images have a relative MSE (normalised with the ground-truth for each modality) of 8.7\% and 3.5\%, respectively.
%Meanwhile, single-modality EIT and QSEI images have relative MSEs of 17\% and 12\%, respectively, showing clear improvements.

\begin{table}[h!]
\small
\begin{tabular}{@{}lcccc@{}}
\toprule
{\textbf{Case}} & \textbf{Single Modality} & \textbf{Single Modality} & \textbf{JTV} & \textbf{JTV} \\ 
 & \textbf{MSE} & \textbf{MSE} & \textbf{MSE} & \textbf{MSE} \\ \midrule
           & EIT                          & QSEI                         & EIT              & QSEI             \\
Single Crack                      & 17                           & 12                           & 8.7              & 3.5              \\
Double Crack                      & 23                           & 19                           & 13             & 10            \\
Through Hole                      & 18                           & 16                           & 8.8              & 6.9              \\
Through Hole (5\%)          & 28                           & 26                           & 19               & 17               \\ 
\end{tabular}
\caption{MSE tabulation for all crack reconstructions using 1\% measurement noise.  The final row reports the MSEs for through hole reconstructions using 5\% added measurement noise. All MSEs are relative and reported as percentages.}
\label{reconinfo}
\end{table}

\begin{figure}[h!]
\centering
\small
\begin{picture}(400,200)
\put(295,90){\includegraphics[width=90pt]{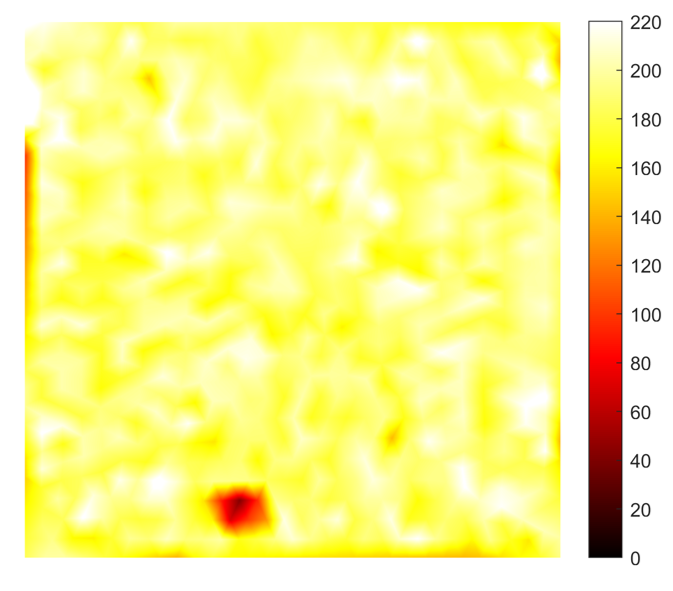}}
\put(200,90){\includegraphics[width=90pt]{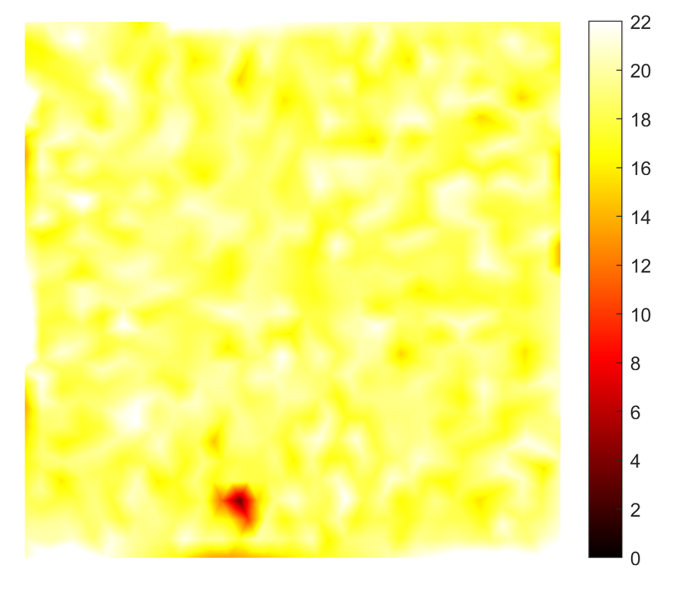}}
\put(105,90){\includegraphics[width=90pt]{images/holeJTVE.png}}
\put(10,90){\includegraphics[width=90pt]{images/holeJTVsig.png}}

\put(10,0){\includegraphics[width=90pt]{images/holesig.png}}
\put(105,0){\includegraphics[width=90pt]{images/holeE.png}}
\put(200,0){\includegraphics[width=90pt]{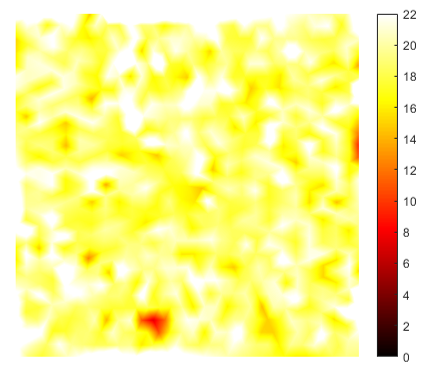}}
\put(295,0){\includegraphics[width=90pt]{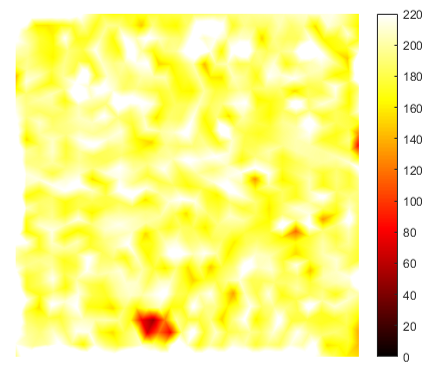}}

\put(0,2){\rotatebox{90}{single-modality}}
\put(0,92){\rotatebox{90}{joint EIT-QSEI}}

\put(40,175){EIT}
\put(135,175){QSEI}
\put(230,175){EIT}
\put(325,175){QSEI}

\put(85,188){1\% noise}
\put(270,188){5\% noise}
\end{picture}

%\textbf{JTV, 1\% noise} & \textbf{JTV, 1\% noise}\\
%\textbf{SM, 1\% noise} & \textbf{SM, 1\% noise}\\

%\textbf{JTV, 5\% noise} & \textbf{JTV, 5\% noise}\\

%\textbf{SM, 5\% noise} & \textbf{SM, 5\% noise}\\

%\textbf{EIT (mS/cm)} & \textbf{QSEI (GPa)} \\
\caption{{Comparison of through hole reconstructions with varying noise level reporting conductivity (mS/cm) and elasticity field $E$ (GPa), with joint total variation and separate single-modality TV. }}
\label{resultsnoisy}
\end{figure}

{What remains to be discussed, however, is the influence of measurement noise on reconstruction quality.
To more closely investigate this, we compare previous through hole  reconstructions (1\% added noise) to substantially corrupted through hole reconstructions (5\% added noise) in Fig. \ref{resultsnoisy}.
Visually, it is immediately apparent that the backgrounds in images reconstructed using JTV presents with less artefacts while the through hole is localized in all cases.
Interestingly, we also note an improvement in sensitivity to the (near zero) conductivity and elasticity values at the through hole location using JTV.
This observation demonstrates a relative improvement in effectiveness in capturing damage processes nearing distinguishability limit by leveraging joint information and regularisation -- in comparison to using a single modality.
Further, we note the quantitative improvements in using JTV, with respect to increasing noise levels, reported in Table \ref{reconinfo}.
}

Based on the visual and quantitative observations, it can be concluded that the JTV reconstructions drastically improved image quality and reduced reconstruction errors.
This may be attributed to a number of key factors, including (a) the increase in spatial information from dual data sets, (b) joint regularisation/penalisation of the EIT and QSEI problems, and (c) the inclusion of simultaneous spatial information in solving the joint problem.
It is interesting to note that, while the JTV problem includes twice as many unknown parameters as the single-modality problems (and is thus, more non-unique/ill-posed), the leveraging of information from items (a-c) was sufficient in significantly improving reconstructions.
This underscores a critical strength of joint inversion techniques, whereby additional prior information can be advantageously included in solving the inverse problem.

In the context of this work, where EIT in relatively insensitive to changes far from the electrodes and QSEI has more uniform sensitivity throughout the domain, EIT reconstruction quality can be drastically improved by mutually-beneficial QSEI measurements. This is supported by a visual improvement of the crack resolution that is especially evident in the EIT reconstruction.
Further yet, joint QSEI image quality is also improved via utilisation of JTV penalisation.

As a whole, however, a primary advantage in using the JTV framework is that we double the amount of information available for assessing structural condition via the simultaneous reconstruction of $\gamma$ and $E$. This leverage will be especially important in the context of missing measurement data, for instance due to detached electrodes, or inaccurate information in the geometry. We expect that a joint framework will be especially beneficial in overcoming the sensitivity of these nonlinear inverse problems to incomplete model information. 

\section{Discussion and outlook}
Nonlinear inverse problems often exhibit a particular sensitivity to measurement noise and incomplete model information. To overcome this problem, we have explored the benefit of joint EIT-QSEI inverse problems by recovering both reconstructions simultaneously. Both problems are of nonlinear nature and hence a combination of the reconstruction task can be naturally realised in a variational formulation. Our results have shown, that both modalities clearly benefit by the auxiliary information provided by the joint total variation used for penalising reconstructions. We employ a weighting strategy in computing a balanced penalty to retain the quantitative information of each modality without over regularising either.

{In this work we are largely motivated by NDT applications.
In future work, we look forward to experimental verification of the joint imaging framework, in particular as applied to simultaneous spatial imaging of coupled sensing skins and DIC patterns.
Such regimes could be used, for example, in detecting distortion induced fatigue cracks in steel bridge girders \cite{dellenbaugh2020development} or characterising tensile hardening in high performance cement-based materials \cite{shen2020influence}.
To this end, as EIT is a penetrative modality, the method could be extended for 3D characterization of large concrete members allowing for simultaneous surface and internal damage monitoring.
Meanwhile, a number of optimisation and parameterisation methods (in particular, nonlinear difference imaging \cite{liu2015nonlinear}) can be adopted to incorporate additional information, enforce physical constraint, and further improve tomographic results.
}

Whereas results are promising, this study can be considered as proof of concept to showcase the synergy between these two nonlinear inverse problems. In particular, even though the combination of two nonlinear inverse problems is unusual, we believe that the presented joint formulation can alleviate the severity of ill-posedness of either problem and improve reconstruction stability in practice. 
Consequently, a more thorough evaluation of improvements in reconstruction quality as well as experimental evaluation of the proposed method will be considered in the future. Nevertheless, the presented results herein are highly promising and should motivate the investigation of possible applications for joint EIT-QSEI imaging, for instance applications in non-destructive evaluation and, likely, in the field of structural health monitoring.
Broadly speaking, the use of fused inverse frameworks for joint imaging remains almost completely unexplored in the aforementioned applications.
Yet, it is the authors' belief that joint imaging can open the door to improved quantitative estimation and reconstruction quality for use in informing the state, condition, and health of engineered materials and structures.

%\enlargethispage{20pt}
\section{Acknowledgements}
\noindent
{\bf Data Accessibility.: } The forward EIT model used is made available via the EIDORS project (\url{http://eidors3d.sourceforge.net/}).  The QSEI forward model used is available at (\url{https://github.com/openQSEI}). \\
\noindent
{\bf Authors' Contributions: } Both authors contributed equally to the design of the study and writing. DS performed the computations. \\
\noindent
{\bf Competing Interests: } The authors declare that they have no competing interests. \\
\noindent
{\bf Funding: } This work was partly funded by Academy of Finland Project 336796 (Finnish Centre of Excellence in Inverse Modelling and Imaging, 2018--2025) as well as Project 334817, and the CMIC-EPSRC platform grant (EP/M020533/1). \\

%\ack{}

%%%%%%%%%% Insert bibliography here %%%%%%%%%%%%%%

\bibliography{bibliography.bib}

\end{document}